\documentclass[12pt,preprint]{aastex}

\newcommand{\teff}{T$_{eff}$}

\newcommand{\wat}{H$_2$O}
\newcommand{\kms}{km s$^{-1}$}

\begin{document}

\title{ On the Nature of the Unique H$\alpha$-Emitting T-Dwarf 
2MASS J12373919+6526148}

\author{James Liebert}
\affil{Steward Observatory, University of Arizona, Tucson AZ 85721, 
liebert@as.arizona.edu}

\author{and}

\author{Adam J.\ Burgasser\altaffilmark{1}}
\affil{Massachusetts Institute of Technology, Kavli Institute for 
Astrophysics and Space Research, 77 Massachusetts Avenue, 
Building 77, Cambridge MA 02139, ajb@mit.edu}

\altaffiltext{1}{Visiting Astronomer at the Infrared Telescope Facility,
which is operated by the University of Hawaii under Cooperative
Agreement NCC 5-538 with the National Aeronautics and Space
Administration, Office of Space Science, Planetary Astronomy Program.}

\begin{abstract}

We explore and discount the hypothesis that the strong, continual
H$\alpha$-emitting T dwarf 2MASS~J12373919+6526148 can be explained as a
young, low gravity, very low mass brown dwarf.  The source is already
known to have a marginally-fainter absolute magnitude than similar T
dwarfs with trigonometric parallax measurements, and has a tangential
velocity consistent with old disk kinematics.  Applying the technique of
Burgasser, Burrows \& Kirkpatrick on new near infrared spectroscopy for
this source, estimates of its {\teff}, $\log{g}$ and metallicity ([M/H])
are obtained.  2M~1237+6526 has a {\teff} $\approx$ 800-850~K.  If [M/H]
is solar, $\log{g}$ is as high as $\sim$5.5 (cgs) and this source is
older than 10 Gyr. We find a more plausible scenario to be a modestly
subsolar metallicity ([M/H] = -0.2) and moderate $\log{g}$~$\sim$~5.0,
implying an age older than 2~Gyr and a mass greater than
0.035~M$_{\sun}$.  The alternative explanation of the unique emission of
this source, involving an interacting, close, double degenerate system,
should be investigated further.  Indeed, there is some evidence of a
{\teff} $<$ 500~K companion to 2M~1237+6526 on the basis of a 
possible $Spitzer~IRAC$ [3.6]--[4.5] color excess.  This excess may, 
however, be caused by a subsolar metallicity.

\end{abstract}


\section{Introduction }

It is well known that the frequency and relative luminosity of H$\alpha$
as a likely measure of chromospheric activity drops rapidly beyond
spectral types M7~V (Kirkpatrick et al. 2000, Gizis et al. 2000, Mohanty
\& Basri 2003, West et al. 2004).  Very few T dwarfs have exhibited this
emission at all, and generally at a very low level compared to active
mid-M dwarfs, as measured by the ratio of L$_{H\alpha}$ / L$_{bol}$
(Burgasser et al.\ 2003).  Occasionally, a vigorous flare is observed in
a late M or L dwarf, causing the ratio to rocket upwards by orders of
magnitude.  For the most part, however, magnetic pressures built up by
the likely dynamo inside these rapidly-rotating objects are not expressed
at the surface, due to the inability of interior plasmas to penetrate the
high resistivities of the predominantly neutral atmospheres (Mohanty
\& Basri 2003).

Two exceptions to this behavior have been found, exhibiting in all
observations levels of H$\alpha$ emission 1-2 orders of magnitude higher
in L$_{H\alpha}$ / L$_{bol}$ than any counterparts of similar spectral
type.  These are the M9.5 dwarf PC~0025+0047 (Schneider et al. 1991) and
the T6.5 dwarf 2MASS~J12373919+6526148 (hereafter 2M~1237+6526,
Burgasser et al.\ 1999).  Continual H$\alpha$ activity of the former
source has been observed for over a decade (Mould et al. 1994;
Mart{\'{\i}}n, Basri, \& Zapatero Osorio 1999).  That the T dwarf seems
to show similar characteristics, though over a shorter time interval so
far, has suggested that emission processes in the two objects may be
related.

Mart{\'{\i}}n et al (1999) made the case that the strong emission lines,
variable amount of optical veiling, and a claimed detection of the
Li~6707\AA\ doublet indicated that PC~0025+0447 is a substellar object
less massive than 0.06M$_{\bigodot}$ and younger than $\sim$1~Gyr.
Their detection of weak K~I and Na~I resonance doublets indicated a
low surface gravity.  They argued, however, that the lack of an infrared
excess and the fact that the H$\alpha$ profile is not similar to
classical T Tauris suggests that chromospheric activity rather than an
active accretion disk is the likely source of the intense emission.  If
the same scenario applies to 2M~1237+6526, this object could be a very
low mass, young, very active brown dwarf.  An age
of 1~Gyr or less implies a maximum mass of 30~M$_J$ (Burrows et
al.\ 2001) for an effective temperature 
{\teff} $\approx$ 800~K based on the Golimowski et al.\
(2004) scale (see their Fig.~6). 

We have explored an alternative hypothesis for this source: that
it is an interacting brown dwarf or double degenerate binary
(Burgasser et al. 2000a, 2002a).  Some possible predictions in the
former paper were tested weakly in the latter paper.  Monitoring of
the J-band flux failed to detect significant variability at the
$\pm$0.025 mag level for periods of hours.  Three separate
measurements of the H$\alpha$ line made at different times with the
Keck/LRIS spectrograph between 1999 and 2001 (over a span of 1.6
yr) were inconclusive in establishing or ruling out variability in the
radial velocity.  Thus, no evidence was uncovered in favor of the
interacting binary hypothesis.  On the other hand, it has not been 
ruled out.

In this paper, we explore the first hypothesis: that 2M~1237+6526
is young, has low gravity, and a very active chromosphere.  We
begin by noting that there is already evidence against this in the
literature (\S~2).  We present additional evidence based on infrared
spectrophotometry in \S~3.  Temperature and gravity sensitive infrared
flux indices are compared with atmospheric models to estimate
T$_{eff}$, $\log{g}$ and the atmospheric abundances ([M/H]) for the
emission line object, as well as age and mass, in \S~4.  
Constraints on the properties of the hypothesized companion to
2M~1237+6526 are discussed in \S~5, based in part on an apparent
mid-infrared color excess detected by Spitzer observations.  Results
are summarized in \S~6.

\section{Luminosity and Kinematics}

Vrba et al. (2004) have published trigonometric parallaxes of 40 L and T
dwarfs which permit the assessment of the absolute magnitude of
2M~1237+6526 in comparison with T dwarfs of similar spectral type.  Its
absolute $J$-band magnitude of 15.88$\pm$0.13 makes it similar to, and
even marginally fainter than, other T6-T7 dwarfs with parallax
measurements.  The same statement applies to absolute magnitudes at $H$
and especially $K_s$ where it is fainter than two T7-7.5 objects.  This
can be seen in Figs.~2--4 of Vrba et al.  It is therefore $\it not$
overluminous compared with other late T dwarfs, as might be expected for
a young brown dwarf with a larger radius (Burrows et al.\ 1997).  This
also rules out the presence of an unseen massive companion; no companion
has yet been detected via high resolution imaging (Burgasser et al.\
2003).

A second argument against youth is the kinematics of this source.  Vrba
et al.\ (2004) measured its tangential velocity to be 56$\pm$3 {\kms},
greater than the median velocity of their sample (39 {\kms}).  In
contrast, PC 0025+0047 has a tangential velocity of only 3.6$\pm$0.4
{\kms} (Dahn et al.\ 2002).  Young objects in the vicinity of the Sun
typically have small relative motions, while older disk and halo objects
can exhibit large velocities (eg. Wielen 1977). A full $UVW$ space
motion analysis of 2M~1237+6526 is not possible as its radial
velocity has not yet been measured.  However, its large tangential speed
does argue against a young age.

\section{The Near Infrared Spectrum}

Far red / near infrared spectrophotometry of 2M~1237+6526 were obtained 
on 8 April 2006 (UT) with the
SpeX spectrograph (Rayner et al. 2003) mounted on the 3-m NASA Infrared
Telescope Facility (IRTF).  Conditions were average, with light cirrus
and moderate seeing (1$\arcsec$ at $J$).  The SpeX prism mode with
a 0$\farcs$5 slit was employed, providing 0.8--2.45 ~$\micron$ spectra
with an average resolution $\lambda/{\Delta}{\lambda}$ $\approx$ 120.
The slit was aligned with the parallactic angle to mitigate color refraction.
Six exposures of 120 s each were obtained at an airmass of 1.50.
The A0~V star HD 99966 was observed immediately afterward for flux calibration
and telluric absorption corrections.  HeNeAr arc lamps and quartz lamp flat
fields were also obtained for dispersion and pixel response calibration.
Data were reduced using the SpeXtool
package version 3.3 (Vacca, Cushing \& Rayner 2003; 
Cushing, Vacca \& Rayner 2004), as
described in detail in Burgasser et al.\ (2004). 

The reduced SpeX spectrum of 2M~1237+6526 is shown in Figure~1, along
with those of the T6.5 dwarfs SDSS~J134646.45-003150.4 (Tsvetanov et
al.\ 2000; hereafter SD 1346-0031) and SDSS~J175805.46+463311.9 (Knapp
et al.\ 2004; hereafter SD 1758+4633), taken from Burgasser, Burrows \&
Kirkpatrick (2006, hereafter BBK).  While the threesome exhibit similar
pseudo-continuum peaks typical of late T dwarfs at 1.27~$\micron$, the
emission line object shows a smaller peak at 1.57~$\micron$ and a very
depressed continuum peak near 2.1~$\micron$ compared with the others.  CH$_4$
and H$_2$O bands are prominent in the $H$ and $K$ bands where the last
two peaks fall.  However, pressure-induced H$_2$ molecular opacity is
very strong at $K$ and affects $H$ as well.  Precisely because of its
strong dependence on pressure, this opacity should be more sensitive
than those of the other two molecules to surface gravity and
metallicity.  The strong dependence of the K band peak on $\log{g}$ is
demonstrated in Fig.~3 of BBK.  A decrease of the atmospheric metal
abundance below solar metallicity also has a similar effect (cf.\ Saumon
et al.\ 1994).

On the other hand, 2M~1237+6526 has the highest Y band
($\sim$1.05$\mu$m) peak of the three sources.  The slope blueward of
this peak is shaped largely by the pressure-broadened red wing of the
0.77 ~$\micron$ \ion{K}{1} doublet.  At higher photospheric pressures
(characteristic of high surface gravity and/or metal-deficient objects),
the slope of this red wing steepens as the near-center wings deepen
(Burrows \& Volobuyev 2003).  Hence, the sharper 1.05 ~$\micron$ peak of
2M~1237+6526 is consistent with the behavior of the $K$-band peak; both
result from either a high surface gravity or lower metallicity.

The \ion{K}{1} and H$_2$ features thus provide evidence that
2M~1237+6526 has the highest surface gravity and/or lowest metallicity
of the three T dwarfs displayed in Figure~1.  This also provides an
explanation for why the M$_K$ value is fainter than the comparison
objects, and suggests that it has among the highest gravities of the
late T dwarfs in the Vrba et al.\ parallax sample.  A higher than
average gravity for a brown dwarf implies a relatively high mass and
older age.  Subsolar metallicity can also imply a higher surface
gravity. Stellar interiors deficient in heavy elements generally have
smaller radii (i.e.\ larger gravity) at a given temperature, and
theoretical calculations show that the same should be true for
substellar entities (Burrows et al. 1993).  Both scenarios imply 
that 2M~1237+6526 is unlikely to be young, and may in fact be quite old.
This hypothesis can be tested using the tools from BBK.

\section{Estimating {\teff} and Surface Gravity Using Spectral Indices
and Constraints on the Mass and Age of 2M~1237+6526}

BBK has demonstrated that a comparison of H$_2$O and color ratio indices
measured on the near infrared spectra of late-type T dwarfs to those
measured on spectral models, after normalizing the latter to the
well-characterized T7.5 brown dwarf companion Gliese 570D (Burgasser et
al.\ 2000b), provides a means of disentangling the effects of {\teff} and
surface gravity for these sources.  These parameters can then be used to
infer age and mass using evolutionary models or a measured bolometric
luminosity.  The technique described in BBK was used to estimate ages
for SD~1346-0031 and SD~1758+4633 of 0.3-0.9 and 1.0-4.9 Gyr,
respectively.  The spectral comparison of Fig.~1 suggests that
2M~1237+6526 is older still.

We implemented the BBK technique for the spectrum of 2M~1237+6526 by
comparing two pairs of indices (H$_2$O-J and K/H, H$_2$O-H and K/J) with
two model sets (Burrows, Sudarsky \& Hubeny 2005; Allard et al.\ 2001).
Figure~2 display the resulting best fit {\teff} and $\log{g}$ phase
spaces for this source, assuming solar metallicity and 10\% uncertainty
in the spectral indices.  While there are slight differences between
these comparisons, all four are consistent with high surface gravity
($\log{g}$ = 5.2--5.5 cgs) and low {\teff} (740--860 K).  The fit using
the Burrows et al.\ models and H$_2$O-J and K/H indices (the nominal set
for BBK) is quite similar to that for 2MASS~J00345157+0523050 (Burgasser
et al.\ 2004), another T6.5 that exhibits strong $K$-band suppression
and a large proper motion, for which BBK estimate an age of
3.4--6.9~Gyr.

As discussed above, subsolar metallicity can also lead to enhanced H$_2$
absorption.  Hence, it is necessary to assess whether the spectrum of
2M~1237+6526 could match that of a metal-poor brown dwarf.  We repeated
the parameter fit analysis using the H$_2$O-J and K/H indices and
subsolar metallicity models from Burrows et al.\ spanning [M/H] = -0.5
to 0.  Results are shown in Figure~3.  As [M/H] is lowered, the best fit
T$_{eff}$ stays around 750-850~K, while the best fit $\log{g}$ generally
decreases.  At [M/H] = -0.5, the best fit $\log{g}$ has dropped a full
dex from the solar abundance fit to 4.5.

There is a clear degeneracy between $\log{g}$ and [M/H] parameters based
on this two-index fitting procedure.  Ideally, one would like to use a
third index, specifically sensitive to metallicity or gravity, to break
the degeneracy.  However, the best candidate for such an index, the Y/J
ratio (ratio between the Y- and J-band pseudocontinuum flux peaks), is
highly sensitive to the \ion{K}{1} line broadening physics, for which
more recent calculations (Burrows \& Volobuyev 2003) are not included in
the Burrows et al.\ subsolar metallicity models (BBK).  As an
alternative, we compared absolute spectral fluxes for 2M~1237+6526 to
those of the Burrows et al.\ models, since a more metal-poor (less
opacity) and lower surface gravity (larger radius) object is overall
more luminous.  Figure~4 compares the absolute flux calibrated spectrum
for 2M~1237+6526 to the best fitting [M/H] = 0 and -0.5 models from the
index analysis.  The data lie between these two extremes and are
inconsistent with them at the 1$\sigma$ flux level, suggesting that
2M~1237+6526 is most likely to be slightly metal-poor ([M/H] $\sim$
-0.2) brown dwarf with a moderate surface gravity ($\log{g}$ $\sim$
5.0).  There are naturally several caveats to consider with this
interpretation, including uncertainties in the {\teff} and $\log{g}$
determinations, systematic errors in the models and the possibility of
excess light from a faint companion.  Nevertheless, the fact that this
object does not exhibit the more pronounced metallicity features of the
peculiar T6 2MASS J09373487+2931409 (Burgasser et al.\ 2002b; hereafter
2M~0937+2931), for which
BBK derive [M/H] = -0.1 to -0.4, argues for a metallicity closer to
solar.

The {\teff} and $\log{g}$ derived for 2M~1237+6526 can be used to infer
the mass and age for this source using the Burrows et al. (1997)
evolutionary models.  For solar metallicity, the four index analyses in
Fig.~1 combined yield an age of 5-10 Gyr and M = 0.043--0.065 M$_{\sun}$.
This is both old and relatively massive for a brown dwarf.  The lowest
metallicity fit ([M/H] = -0.5) yields a significantly younger age
($\sim$0.4 Gyr) and lower mass ($\sim$0.015 M$_{\sun}$), but for the
reasons discussed above this interpretation is not favored.  Rather, for
$\log{g} \gtrsim 5.0$ and the best fit {\teff}s (800--850~K), 2M~1237+6526
is likely to be older than 2~Gyr and more massive than 0.035~M$_{\sun}$.
This conclusion is supported by the deduced mass of this object from the
estimated {\teff} and $\log{g}$ values and the measured bolometric
luminosity (Vrba et al. (2004), which yield M =
$Lg/4{\pi}G{\sigma}T^4_{eff}$ $\approx$ 0.04 M$_{\sun}$, albeit with
significant uncertainty.

\section{Constraints on the Hypothetical Companion of 2M~1237+6526: 
Has It Been Detected?}


If 2M~1237+6526 has a companion overfilling its Roche lobe, we concluded
in Burgasser et al. (2000) that this companion's mass must be $<$63\%
that of the primary.  Hence, for a primary with mass 0.04~M$_{\sun}$
(0.065~M$_{\sun}$) at an age of 2~Gyr (10~Gyr), then the companion must
have a mass $<$0.025~M$_{\sun}$ ($<$0.04~M$_{\sun}$) and {\teff}$<$650~K
($<$550~K).  The hypothesized semi-detached companion, under these
conditions, must be substantially cooler than its primary, and also
cooler than any currently-known T dwarf.

Could such a companion have been detected in $Spitzer - IRAC$
3.6--8~$\micron$ observations of 2M~1237+6526?  We show in Figure~5 a
color magnitude diagram for the two most sensitive $IRAC$ bands, the
absolute 3.6~$\micron$ magnitude vs.\ the [3.6]--[4.5] color for objects
T5 and later with trigonometric parallaxes, including two known mid-T
binaries with components of similar brightness.  The data are taken from
Patten et al.~(2006), plus values for the T7.5 companion to HD~3651 are
taken from Luhman et al.~(2006).  Also shown are synthetic photometry
based on the models similar to those of Burrows, Sudarsky \& Lunine
(2003).  As discussed in Patten et al.\ (2006), the disagreement between
the models and data in this plot likely arises from overestimated
4.5~$\micron$ fluxes in the models.  This is likely due to
underestimated abundances of CO, that appears to be enhanced by vertical
upwelling (Noll, Geballe, \& Marley 1997; Oppenheimer et al.\ 1998;
Saumon et al.\ 2000).  Absorption due to this molecule's fundamental
band at $\sim$4.67~$\micron$ probably causes the observed fluxes to be
weaker than the models predict, and therefore the [3.6]--[4.5] colors to
be bluer.

Trends in the data suggest a blue envelope of [3.6]--[4.5] colors that
parallels similar trends in the models.  This envelope, which we
approximate as
\begin{equation}
[3.6]-[4.5] \geq -7.05 + 0.58M_{[3.6]}
\end{equation}
over the range 14.0 $<$ $M_{[3.6]}$ $<$ 15.3, should trace out the
oldest and most massive single brown dwarfs, as the models predict that
lower surface gravities (i.e., younger ages) yield redder colors.  Most
of the T5--T8 dwarfs from this sample lie close to the envelope (with
small color dispersion, although there is some spread in $M_{[3.6]}$
along the envelope amongst objects in a given spectral subclass,
possibly due to classification uncertainties).  2M~1237+6526, on the
other hand, sits modestly above or redward of this envelope.  As it is
an apparently old source, this excess suggests the presence of a faint,
unresolved companion.

By constraining the primary to lie on the
the envelope line ($M_{[3.6]} \approx$ 14.4, [3.6]--[4.5] $\approx$ 1.3)
-- marked by a green circle -- we show the effect of
adding a companion with T$_{eff}$= 500~K, log g = 5.0
and solar composition, based on theoretical photometry
calculated for us by A. Burrows.  This companion has 
$M_{[3.6]}$= 17.28, [3.6]--[4.5] = 3.12.  Coadding the
fluxes in each of the two bands places the combined-light
binary at the position of the red circle ($M_{[3.6]}$= 14.33,
[3.6]--[4.5] = 1.58).  Thus, if our speculation is correct that
2M~1237+6526 has a cooler companion, it is probably more than 300~K
cooler than its primary, which is consistent with the constraints
placed on the companion by the interacting binary scenario.

Note also that both the T6p 2M~0937+2931 and the T6.5 2MASS
J10475385+2124234 (Burgasser et al.\ 1999; hereafter 2M~1047+2124; not
plotted in the figure as it has values nearly identical to 2M~1237+6526)
also lie redward of the proposed envelope.  While it is possible that
the latter source is young and has a low surface gravity, the former
source has been shown to be an old, high surface gravity brown dwarf
with subsolar metallicity (BBK).  Since 2M~1237+6526 also appears to be
slightly metal-deficient, its red [3.6]--[4.5] color may instead be a
metallicity effect.  Lower heavy element abundances would increase the
atmospheric pressures, which might enhance the formation of CH$_4$ while
weakening CO.  The latter could enhance the 4.5~$\mu$m flux, while the
former could weaken the 3.6~$\mu$m flux.  Thus, the $IRAC$ colors of
2M~1237+6526 cannot unambiguously confirm the presence of a low-mass
unresolved companion.

\section{Summary and Implications}

On the basis of our spectral analysis, as well as the object's 
luminosity and kinematics, we conclude that 2M~1237+6526 is likely to be
a rather old, massive and slightly metal-poor brown dwarf.  Youthful
activity, which is cited as the source for the emission properties of
PC~0025+0047, cannot explain this T dwarf's persistent H$\alpha$
emission.

Other possibilities, including accretion from a close companion, are
worth further investigation.  The ``tests'' of the double degenerate
hypothesis involving a low rate of transfer have not been conclusive.
The problem with searching for possible radial velocity variations was
described in the Introduction.  The absence of large infrared variations
sought in the Burgasser et al. (2002a) study rules out only a limited
phase space for the interacting binary model.  The $IRAC$ photometry
suggests a [3.6]--[4.5] color excess which could result from the
presence of an unresolved cool companion, but the excess may also be
attributable to a lower metallicity.  Any companion would have to be
appreciably lower in mass, and probably at least 300~K cooler.
Generally speaking, as model atmospheres are improved to predict more
accurate 4.5~$\mu$m fluxes, including the dependence on metallicity,
searching for an excess in this band may result in the detection of
unresolved companions later in type than any currently-known T dwarf.

\acknowledgements

We thank our telescope operator Eric Volquardsen and instrument
specialist John Rayner, for their support during the SpeX observations.
Helpful discussions with Michael Cushing are acknowledged, and we thank
him for help with the IRAC magnitude system.  We also thank Adam Burrows
for providing additional theoretical $Spitzer~IRAC$ photometry for our
analysis.  Our referee, Gibor Basri, provided a useful critique of our
original manuscript that greatly improved it.  This publication makes
use of data products from the Two Micron All Sky Survey, a joint project
of the University of Massachusetts and the Infrared Processing and
Analysis Center/California Institute of Technology, funded by the
National Aerospeace and Space Administration and the National Science
Foundation.  The authors wish to extend special thanks to those of
Hawaiian ancestry on whose sacred mountain we are privileged to be
guests.

\clearpage

\begin{figure}
\epsscale{0.8}
\plotone{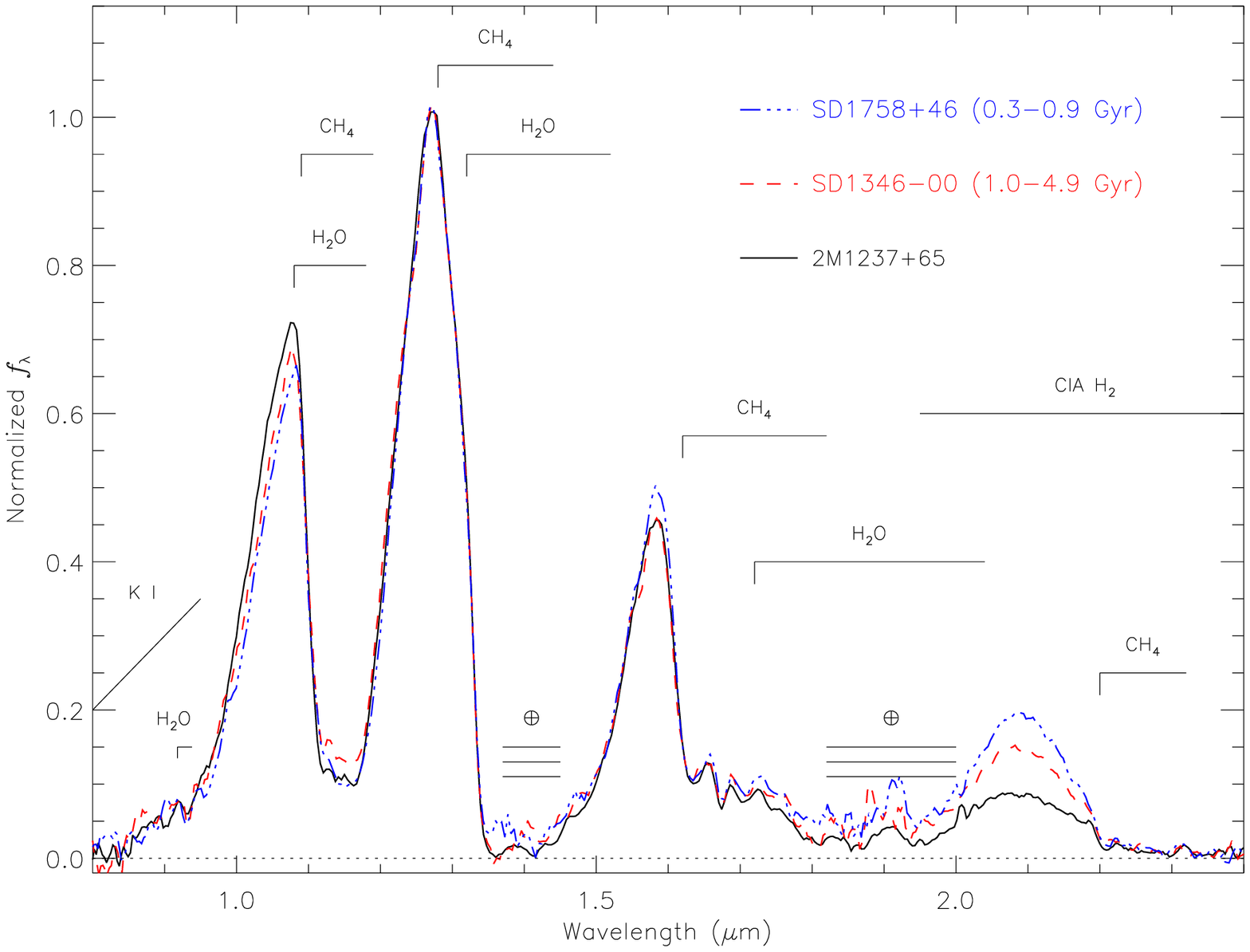}
\caption{SpeX spectra of the T6.5 dwarfs 2M~1237+6526 (black solid line), 
SD~1346-0031 (red dashed line) and SD~1758+4633 (blue dot dashed line).
All data are normalized at 1.27 ~$\micron$.  Major spectral features are
labelled, as well as telluric absorption bands ($\oplus$ symbols).  The
spectra are largely identical, with the exception of significant
differences in the peak $K$-band flux and slight differences in the
1.05~~$\micron$ $Y$-band and 1.6~~$\micron$ $H$-band peaks.  Age estimates
from BBK for SD~1346-0031 and SD~1758+4633 are listed.
\label{fig_spectra}}
\end{figure}

\clearpage

\begin{figure}
\includegraphics[width=0.45\textwidth]{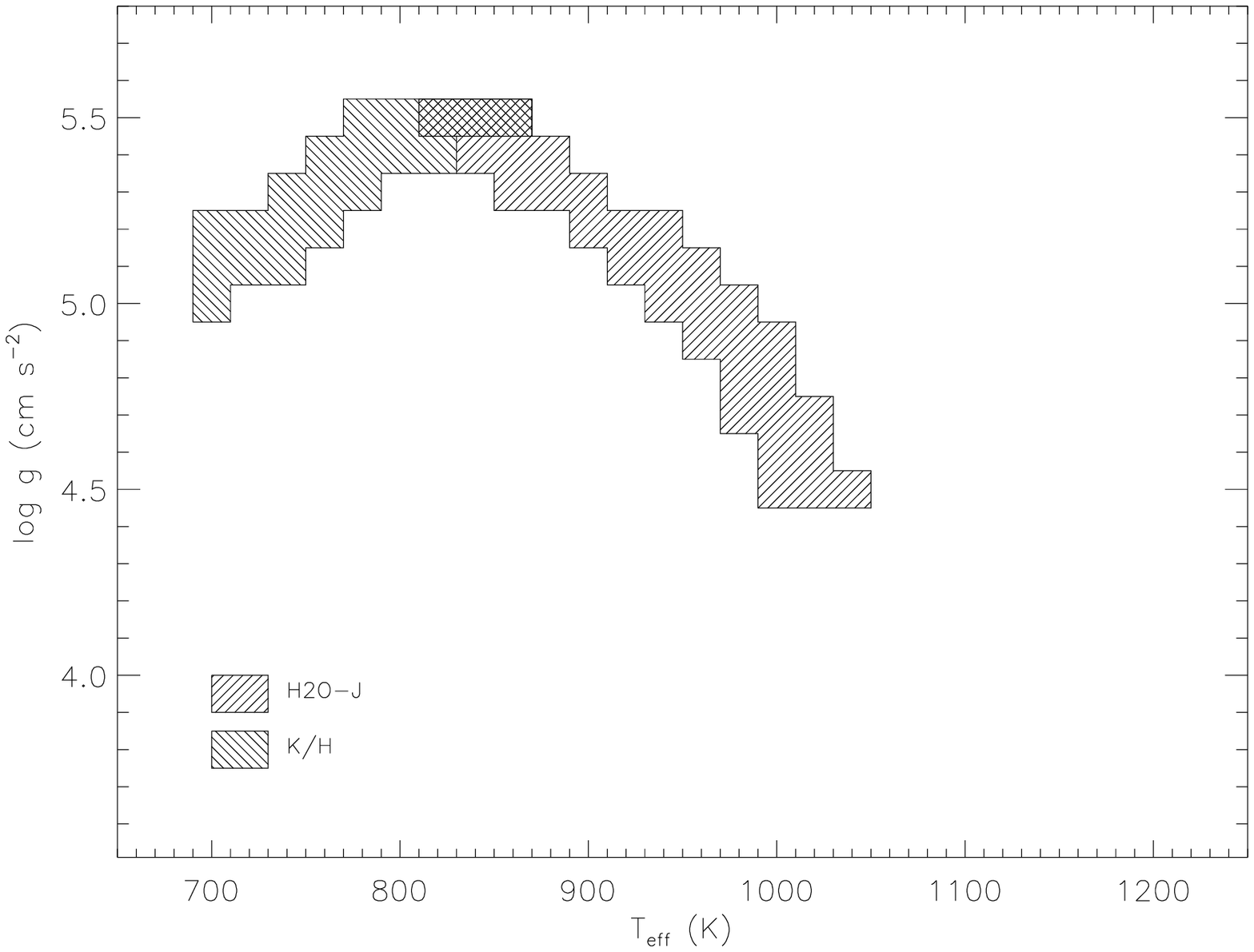}
\includegraphics[width=0.45\textwidth]{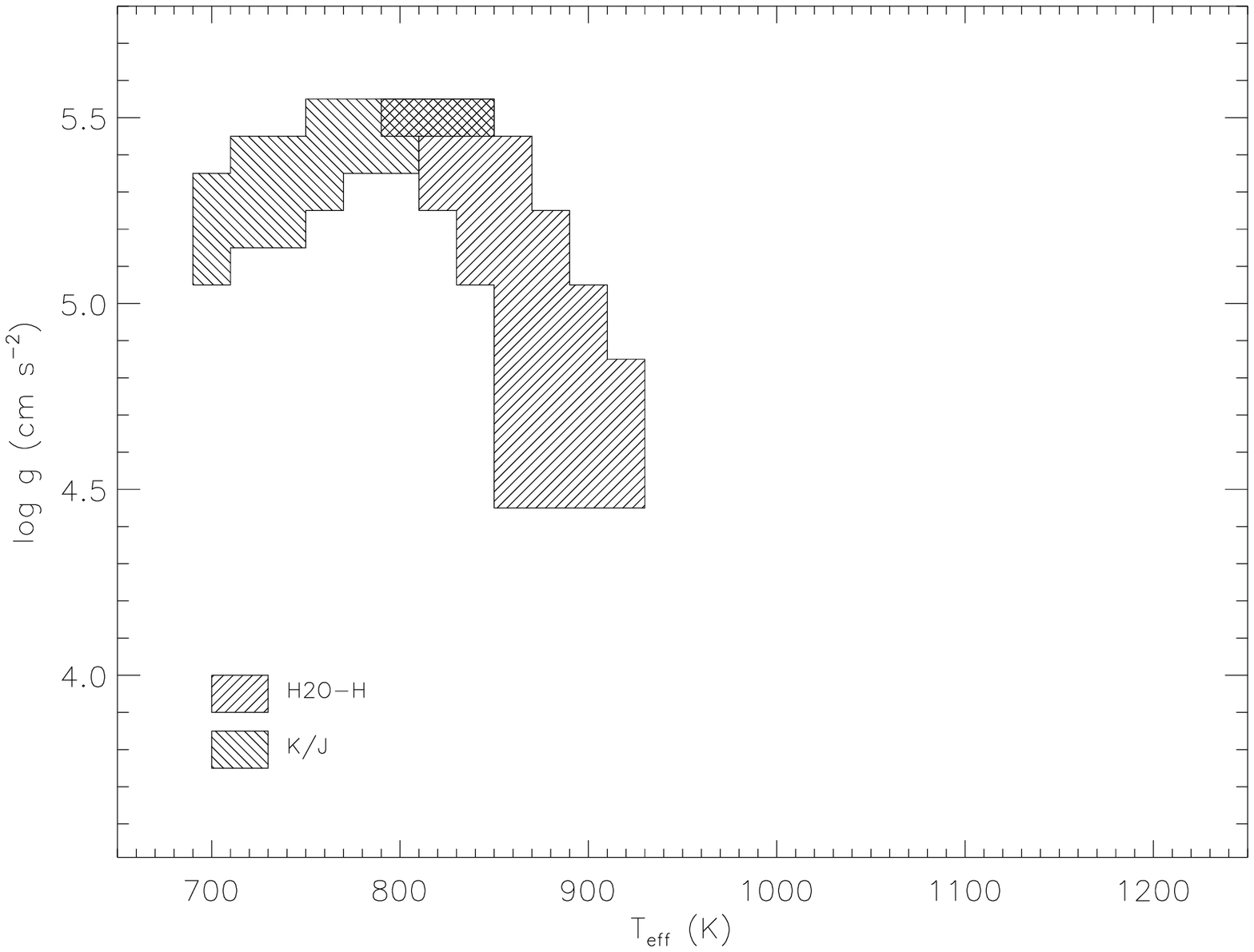}
\includegraphics[width=0.45\textwidth]{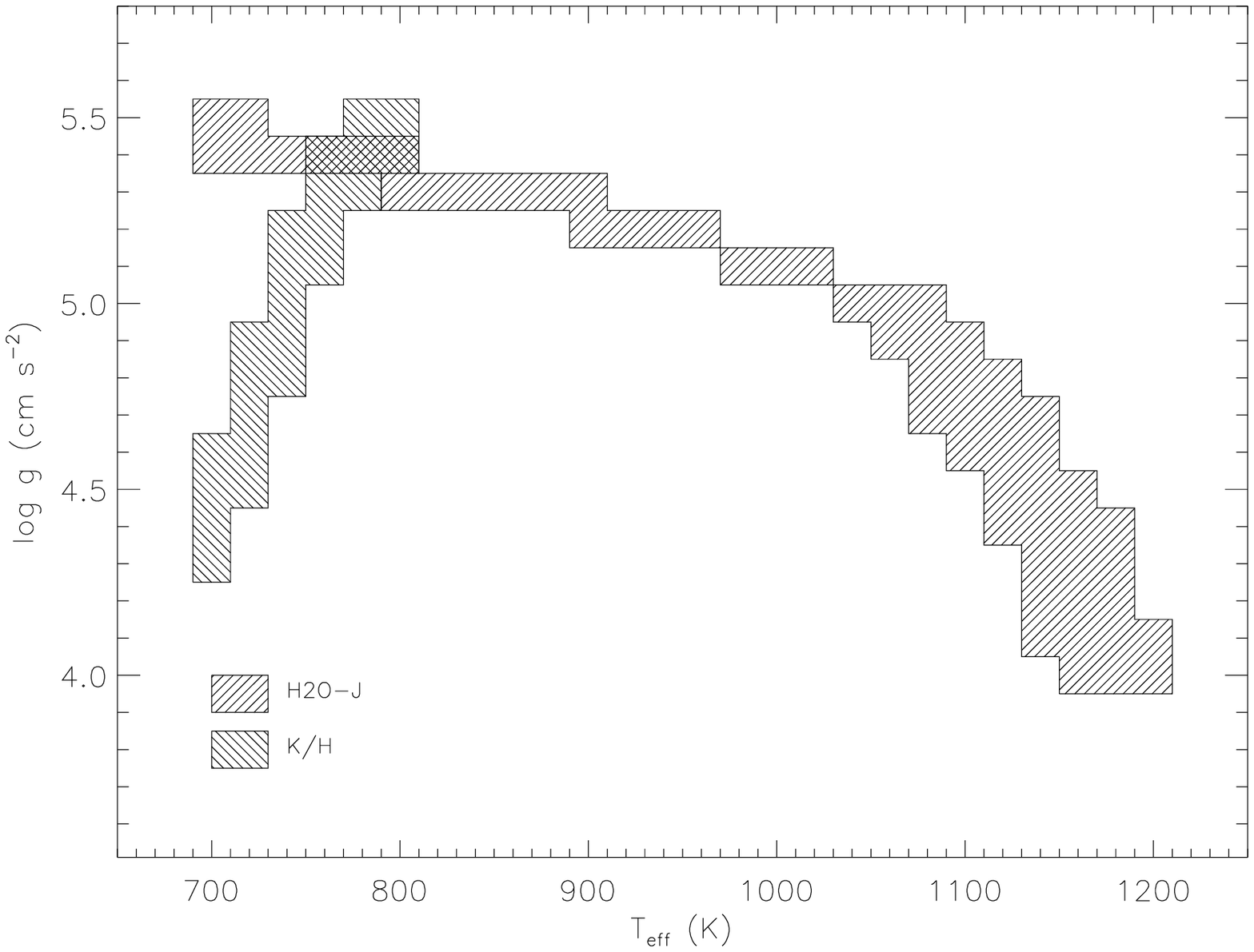}
\includegraphics[width=0.45\textwidth]{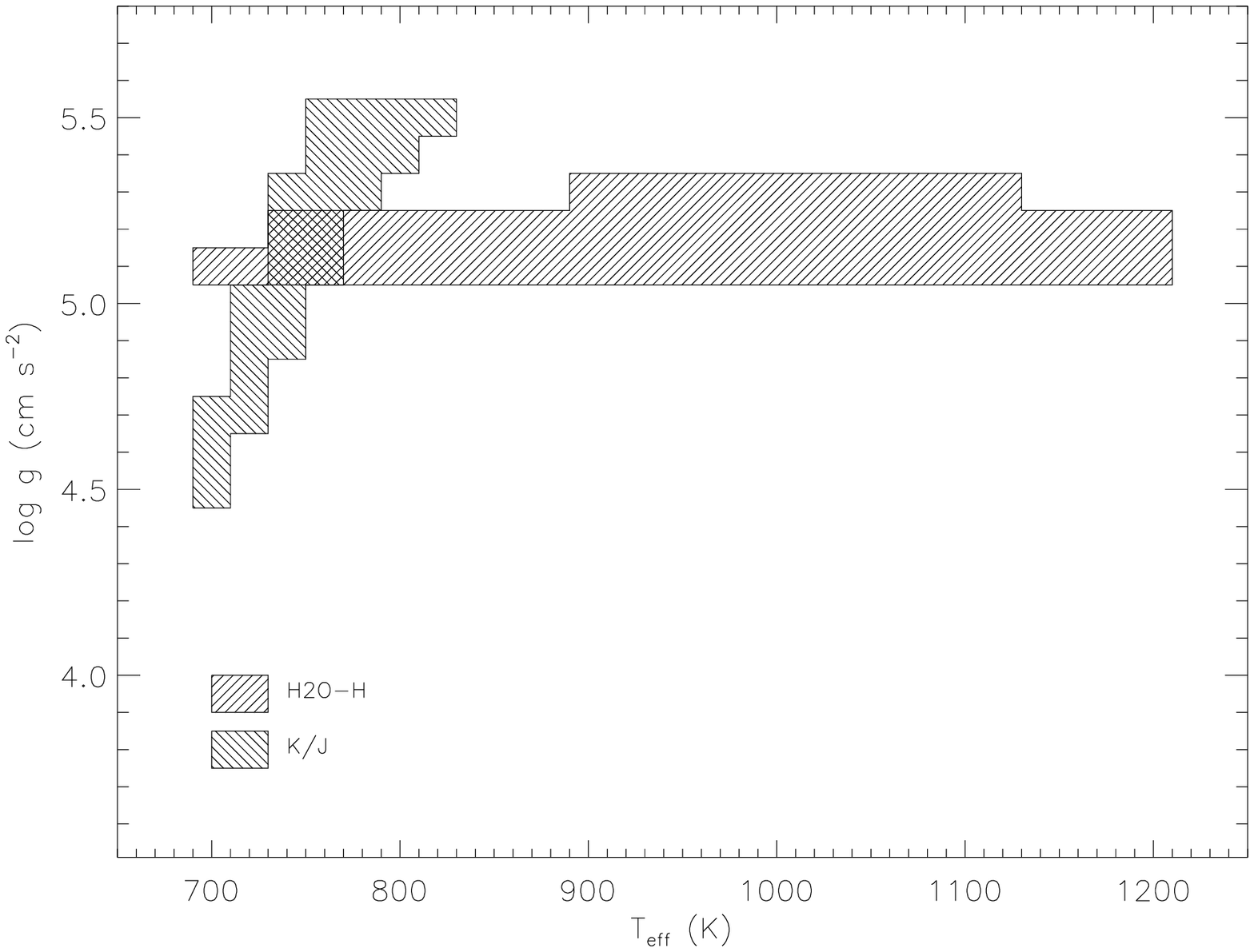}
\caption{T$_{eff}$ and gravity constraints for 2M~1237+6526 based on
the method of BBK. The top panels show index comparisons to the models
of Burrows, Sudarsky \& Hubeny (2005), the bottom panels show
comparisons to the COND models of Allard et al.\ (2001).  The left
panels compare {\wat}-J and K/H indices, while the right panels compare
{\wat}-H and K/J indices.  All four comparisons yield similar results,
although the {\wat}-H and K/J comparison to the COND models yield
somewhat lower {\teff}s and gravities.
\label{fig_tg}}
\end{figure}

\clearpage

\begin{figure}
\epsscale{0.8}
\plotone{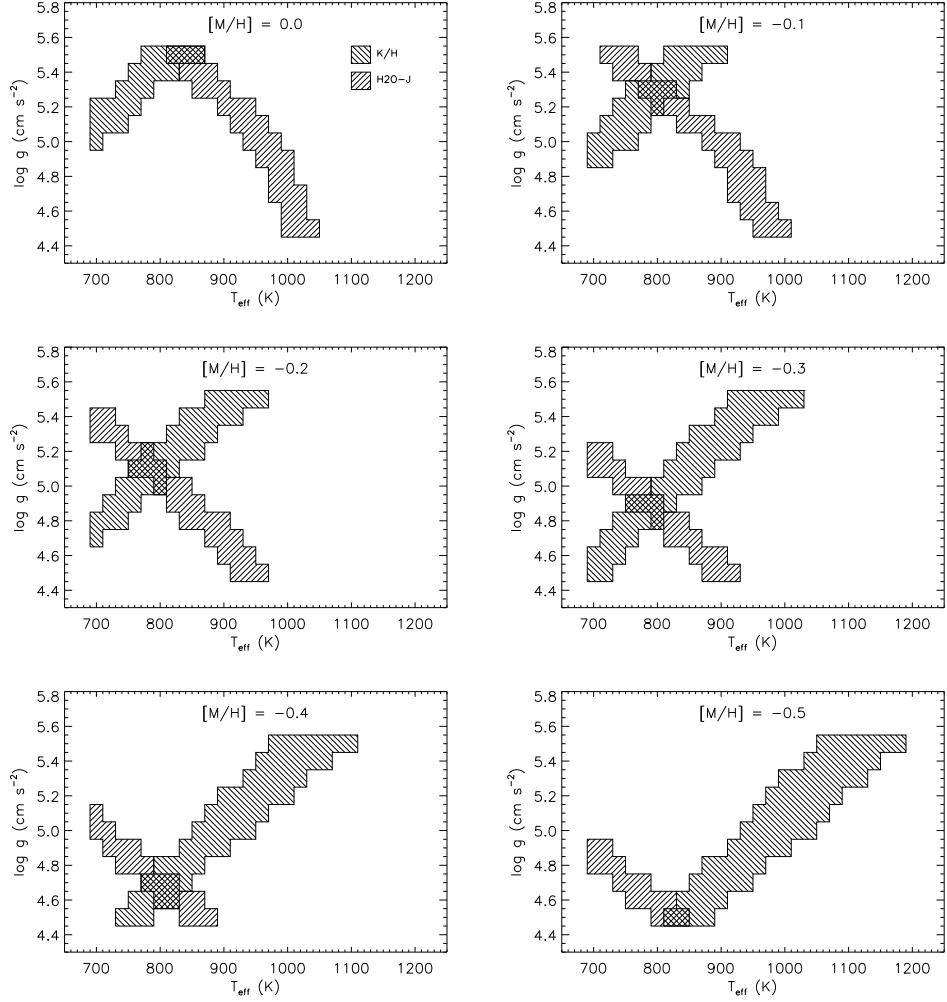}
\caption{Same as Figure~\ref{fig_tg}, but comparing {\wat}-J and K/H
indices for subsolar metallicity models from Burrows, Sudarsky \& Hubeny
(2005) spanning [M/H] = 0 to -0.5.  Note that lower metallicity models
permit lower surface gravities, and hence lower mass and younger age
constraints for this pair of indices.
\label{fig_tgz}}
\end{figure}

\clearpage

\begin{figure}
\epsscale{0.8}
\plotone{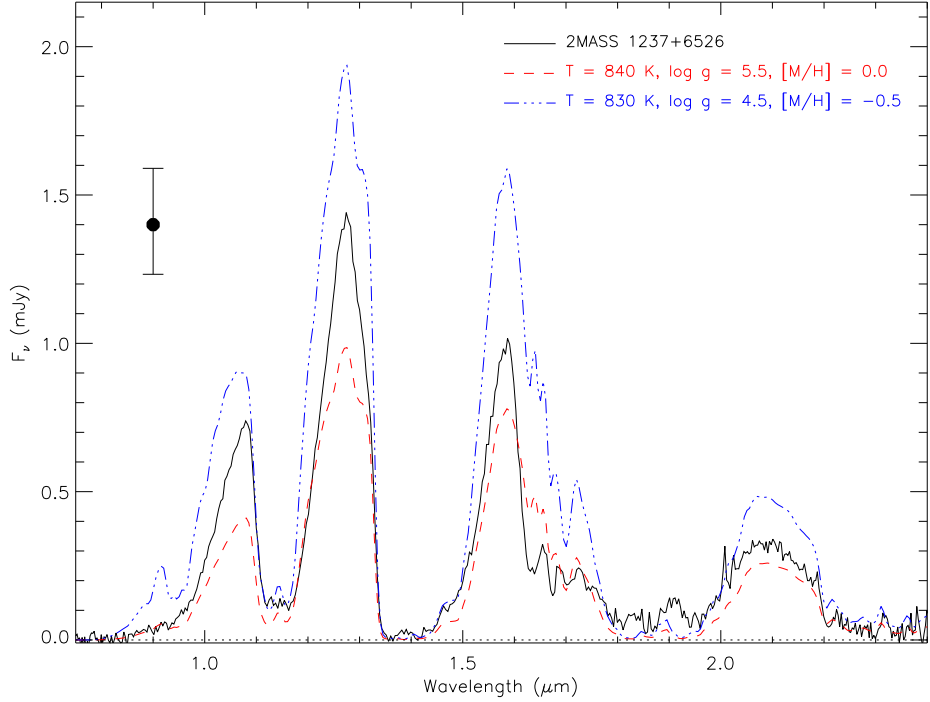}
\caption{Comparison of absolute spectral fluxes ($F_{\nu}$)
for 2M~1237+6526 (black solid line) to theoretical models for the best
fitting [M/H] = 0 (red dashed line; {\teff} = 840~K and $\log{g}$ = 5.5)
and [M/H] = -0.5 (blue dot-dashed line; {\teff} = 830~K, $\log{g}$ =
4.5) parameters.  Uncertainty in the absolute calibration of 2M
1237+6526, based on $J$-band photometry and parallax measurements from
Vrba et al.\ (2004), is indicated by the error bar in the upper left
corner of the plot.  The spectral data appear to favor an intermediate
metallicity ([M/H] $\sim$ -0.2) and surface gravity ($\log{g} \gtrsim
5.0$) for this source, assuming no significant contribution from an
unseen companion.
\label{fig_model}}
\end{figure}

\clearpage

\begin{figure}
\epsscale{0.8}
\plotone{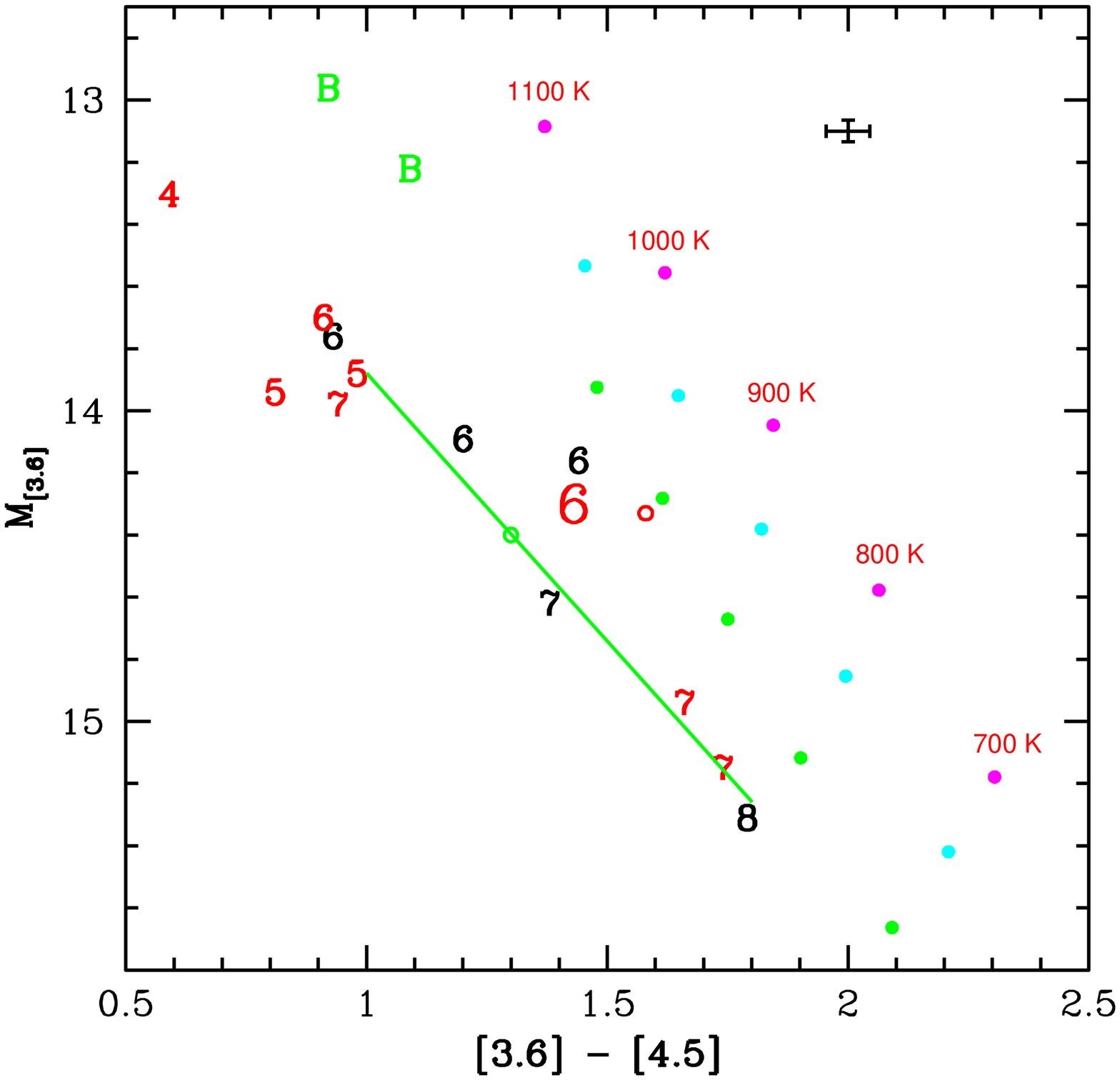}
\caption{The M$_{[3.6]}$ vs. [3.6]--[4.5] diagram for T dwarfs later
than T4.0 with trigonometric parallaxes, observed in Patten et al.
(2006) and Luhman et al.~(2006).  Each spectral class is plotted as a
black numeral if the subtype is exact, or as a red color if there is 0.5
subtype to be added.  2M~1237+6526 is shown as the enlarged red ``6''
(for T6.5).  To avoid confused symbols, 2M~1047+2124 with virtually
identical values to 2M~1237+6526 is not plotted.  Photometry for two
known, equal-brightness T dwarf binaries (2M15344984-2952274 and
2M12255432-2739466, Burgasser et al. 2003) are indicated by the ``B''
symbols.  Filled circles are synthetic photometry based on the models of
Burrows, Sudarsky \& Lunine (2003) as reported in Patten et al.\ (2006)
for \teff\ as labelled, for $\log{g}$ values of (from right to left in
cgs units) 4.5 (magenta), 5.0 (cyan), and 5.5 (green).  The solid line
delineates our proposed ``envelope'' of [3.6]--[4.5] colors for 14.0 $<$
$M_{[3.6]}$ $<$ 15.3, as suggested by the data (Equation 1).  According
to the models, this envelope should correspond to the oldest, single
brown dwarfs in the sample.  The green and red open circles indicate the
estimated position of a high gravity T6.5 and show the effect of adding
a T$_{eff}$ = 500~K companion to that source, as discussed in the text.
\label{fig_comp}}
\end{figure} 


\begin{references}

\reference{} Allard, F., Hauschildt, P.\ H.,
Alexander, D.\ R., Tamanai, A., \&  Schweitzer, A. 2001, \apj, 556, 357


\reference{} Burgasser, A.\ J., Burrows, A., \& Kirkpatrick, J.\ D. 
2006, \apj, 639, 1095 (BBK)

\reference{} Burgasser, A.\ J., Kirkpatrick, J.\ D.,
Liebert, J., \& Burrows, A. 2003, \apj, 594, 510

\reference{} Burgasser, A.\ J., Kirkpatrick, J.\ D., Reid, I.\ N., 
Brown, M.\ E., Miskey, C.\ L., \& Gizis, J.\ E. 2003, \apj, 586, 512

\reference{} Burgasser, A.\ J., Kirkpatrick, J.\ D.,
Reid, I.\ N., Liebert, J., Gizis, J.\ E., \& Brown, M.\ E. 2000a, \aj, 120,
473

\reference{} Burgasser, A.J., Liebert, J., Kirkpatrick, J.D., \& 
Gizis, J.E. 2002a, \aj, 123, 2744.

\reference{} Burgasser, A.J., McElwain, M.W., Kirkpatrick, J.D., Cruz,
K.L., Tinney, C.G., \& Reid, I.N. 2004, \aj, 127, 2856

\reference{} Burgasser, A.\ J., et al.  1999, \apj, 522, L65

\reference{} Burgasser, A.\ J. et al. 2000b, \apj, 531, L57  

\reference{} Burgasser, A.\ J., et al. 2002b, \apj, 564, 421

\reference{} Burrows, A., Hubbard, W.B., Lunine, J.I., \& Liebert, J. 
2001, {\it Rev. Mod. Phys.}, 73, 719 

\reference{} Burrows, A., Hubbard, W.B, Saumon, D., \& Lunine,
J.I. 1993, \apj, 406, 158 

\reference{} Burrows, A., Sudarsky, D., \& Lunine, J.\ I. 2003, \apj,
596, 587

\reference{} Burrows, A., \& Volobuyev, M. 2003, \apj,
583, 985

\reference{} Burrows, A., et al.\ 1997, \apj, 491, 856


\reference{} Cushing, M.C., Vacca, W.D., \& Rayner, J.T. 2004, \pasp, 
116, 362

\reference{} Dahn, C.\ C., et al. 2002, \aj, 124, 1170

\reference{} Golimowski, D., et al. 2004, \aj, 127, 3516

\reference{} Gizis, J.E., Monet, D.G., Reid, I.N., Kirkpatrick, J.D.,
Liebert, J. \& Williams, R.J. 2000, \aj, 120, 1085

\reference{} Kirkpatrick, J.D., Reid, I.N., Liebert, J., Gizis, J.E.,
Burgasser, A.J., Monet, D.G., Dahn, C.C., Nelson, B., \& Williams,
R.J. 2000, \aj, 120, 447

\reference{} Knapp, G.\ R., et al. 2004, \aj, 127, 3553. 


\reference{} Luhman, K.L. et al. 2006, \apj, in press 

\reference{} Mart{\'{\i}}n, E.L., Basri, G., \& Zapatero Osorio,
M.R. 1999, \aj, 118, 1005

\reference{} Mohanty, S. \& Basri, G. 2003, \apj, 583, 451 

\reference{} Mould, J., Cohen, J., Oke, J.B., \& Reid, I.N. 1994, 
\aj, 107, 222 

\reference{} Noll, K.\ S., Geballe, T.\ R., \&
Marley, M.\ S. 1997, \apj, 489, L87

\reference{} Oppenheimer, B.\ R., Kulkarni, S.\ R.,
Matthews, K., van Kerkwijk, M.\ H. 1998, \apj, 502, 932

\reference{} Patten, B.M. et al. 2006, \apj, in press (astro-ph/0607537)

\reference{} Rayner, J.T., Toomey, D.W., Onaka, P.M., Denault, A.J., 
Stahlberger, W.E., Vacc, W.D., Cushing, M.C., \& Wang, S. 2003, \pasp, 
115, 362

\reference{} Saumon, D., Bergeron, P., Lunine, J.\ I.,
Hubbard, W.\ B., \& Burrows, A. 1994, \apj, 424, 333

\reference{} Saumon, D., Geballe, T.R., Leggett, S.K., Marley, M.S., 
Freedman, R.S., Lodders, K., Fegley, B., Jr., \& Sengupta, S.K. 2000, 
\apj, 541, 374 

\reference{} Schneider, D.P., Greenstein, J.L., Schmidt, M. \& Gunn, J.E. 
1991, \aj, 102, 1180

\reference{} Tsvetanov, Z.\ I., et al. 2000, \apj,
531, L61

\reference{} Vacca, W.D., Cushing, M.C., \& Rayner, J.T. 2003, \pasp, 
115, 389 

\reference{} Vrba, F.J. et al. 2004, \aj, 127, 2948

\reference{} West, A.A. et al. 2004, \aj, 128, 426

\reference{} Wielen, R. 1977, \aap, 60, 263

\end{references}
\end{document}